\documentclass[twocolumn]{article}

\begin{document}

\title{\bfseries
        \vspace*{-0.3in}%
\begin{flushright}\large AIAA-98-3142
\end{flushright}        \vspace*{0.1in}{\large \textbf{ON THE POSSIBILITY OF
A PROPULSION DRIVE CREATION THROUGH A LOCAL MANIPULATION OF SPACETIME
GEOMETRY}} }
\author{Vesselin Petkov \\
Physics Department, Concordia University\\
1455 De Maisonneuve Boulevard West\\
Montreal, Quebec, Canada H3G 1M8\\
E-mail: vpetkov@alcor.concordia.ca}
\date{}
\maketitle





\begin{center}
\textbf{{\underline{ABSTRACT}} }
\end{center}

Since the shape of a free body's worldline is determined by the geometry of
spacetime a local change of spacetime geometry will affect a body's
worldline, i.e. a body's state of motion. The exploration of this
possibility constitutes a radically new approach to the idea of how a body
can be propelled: instead of applying a force to the body itself, the
geometry of spacetime is subjected to a local manipulation which in turn
results in the body's motion.

\medskip

\begin{center}
\textbf{{\underline{INTRODUCTION}}\\[0pt]
}
\end{center}

It is the geometry of spacetime that determines the shape of a free body's
worldline. If spacetime is flat, a straight worldline represents a body
moving by inertia - the body is moving with a constant velocity and its
motion is non-resistant. If spacetime is curved, a non-resistant (inertial)
motion is represented by a geodesic worldline. In this case a body moving
along a geodesic worldline (for instance, a falling body) offers no
resistance to moving with acceleration with respect to a given reference
frame (the Earth's surface, for example). It follows from here that if we
can change the geometry of spacetime locally, we can change the shape of a
body's geodesic worldline which is equivalent to making a body move without
subjecting it to any force. Such a force-free motion is non-resistant no
matter whether it is accelerated since the body continues to move along its
geodesic worldline which has a different shape in the locally changed
spacetime geometry.

The essential information concerning spacetime geometry is given by Riemann
curvature tensor - if it is zero the spacetime is flat, otherwise it is
curved. The nature of spacetime curvature, however, has been an unsolved
puzzle in physics. In this paper an approach based on the classical
electromagnetic mass theory which provides an insight into what the
curvature of spacetime may mean is outlined.

One of the consequences of general relativity is that the velocity of
electromagnetic signals (or simply the velocity of light) in the vicinity of
massive objects is anisotropic; it is believed that this anisotropy is
caused by the spacetime curvature. Using the anisotropic speed of light in
the calculation of the self-force with which each non-inertial elementary
charged particle (an electron, for example) acts upon itself on account of
its own electric field leads to the following consequences. Due to the
anisotropy of the speed of light the electric field of an electron on the
Earth's surface is distorted which gives rise to a self-force originating
from the interaction of the electron's charge with its distorted electric
field. This self-force tries to force the electron to move downwards and
coincides with what is traditionally called a gravitational force. The
electric self-force is proportional to the gravitational acceleration $%
\mathbf{g}$ and the coefficient of proportionality is the mass ''attached''
to the electron's electric field which proves to be equal to the electron's
mass. In such a way the electron's passive gravitational mass turns out to
be purely electromagnetic in origin. This means that the only intrinsic
property of an electron is its charge (and the resulting electromagnetic
field). The mass of an electron is a secondary property corresponding to the
energy stored in the electron's electromagnetic field. Simply put, there is
no mass; there are only charges and electromagnetic fields.

The anisotropy of the speed of light is compensated if an electron is
falling toward the Earth's surface with an acceleration \textbf{g}. In other
words, the electron is falling in order to keep its electric field not
distorted. A Coulomb (not distorted) field does not give rise to any
self-force acting on the electron; that is why the motion of a falling
electron is non-resistant as general relativity predicts. If an electron is
prevented from falling it can no longer compensate the anisotropy of
spacetime, its field distorts and as a result a self-force pulling the
electron downwards arises.

Therefore the anisotropic speed of light around objects and the
electromagnetic mass approach fully account for the gravitational properties
of charged particles. In fact, this approach fully explains the
gravitational attraction between bodies as well since the constituents of
the neutron are also charged sub-particles (quarks). This is an indication
that the anisotropy of spacetime (manifesting itself in the anisotropy in
the propagation of electromagnetic signals) is not caused by a curvature of
spacetime (since no curvature hypothesis is necessary) but itself can be
interpreted as a curvature or as a gravitational field. In such a way, it
turns out that Riemann curvature tensor, in fact, describes the spacetime
anisotropy.

\renewcommand{\thefootnote}{\fnsymbol{footnote}} As an electron's passive
gravitational mass in this approach is electromagnetic, its active
gravitational mass, being equal to its passive gravitational mass, is
electromagnetic too. And since it is the active gravitational mass of an
electron that produces its gravitational field, i.e. the anisotropy of
spacetime around the electron, it follows that the spacetime anisotropy
originates from the electron's charge and electric field (since an electron
possesses only charge and electric field). This means that the anisotropy of
spacetime and therefore the spacetime geometry itself are locally \emph{in
principle} controllable since it is the charge and the electric field of an
electron that cause the anisotropy of spacetime in its vicinity\footnote{%
The controllability of spacetime geometry is a direct consequence of the
fact that the sources of spacetime anisotropy (charges and electric fields),
being electromagnetic phenomena themselves, are in principle controllable.}.

The behaviour of an electron in an accelerated reference frame is identical
to that of an electron in the Earth's gravitational field (the anisotropy in
the speed of light in this case is caused by the frame's accelerated
motion). The electromagnetic field of an accelerated electron is distorted
which results in an electromagnetic self-force acting upon the electron and
resisting its accelerated motion. It is proportional to the electron's
acceleration, the coefficient of proportionality being exactly its
electromagnetic mass$^{1-10}$ (i.e. the mass ''attached'' to its
electromagnetic field). In such a way, like gravitation and the passive
gravitational mass, inertia and the inertial mass of an electron also turn
out to be electromagnetic in nature\footnote{%
What part of the mass is electromagnetic is an open question now. It is an
accepted fact, however, that at least part of the mass of every charged
particle is electromagnetic in origin. It has not been realized so far that
an immediate consequence of this fact is that both inertia and gravitation
prove to be at least partly electromagnetic as well.}. And if we can control
other electromagnetic phenomena nothing in principle prevents us from doing
so to inertia and gravitation as well.

The approach followed in this paper leads to and confirms the basic results
of recent publications by B. Haisch, A. Rueda and H. Puthoff$^{11-14}$
regarding the electromagnetic nature of inertia and gravitation. In their
view inertia and gravitation result from interactions between the
electromagnetic zero-point field and the elementary charged particles of
matter. The fact that Haisch, Rueda and Puthoff's zero-point field approach
and the source-field approach of this paper, which are in fact
complimentary, come up with the same interpretation of inertia and
gravitation can hardly be a pure coincidence.

In the sections which follow the anisotropic velocity of light in a
gravitational field is calculated and it is shown that taking it into
account in the calculation of the electric field of an electron in a
gravitational field fully accounts for the electron's gravitational
properties. The calculation of the electric potential, electric field and
the self-force of a non-inertial electron is non-covariant since the physics
is more transparent in this case; a covariant formulation is easily
obtainable$^{10}$. At this stage it appears that quantum mechanical
treatment of the electromagnetic mass is not possible since quantum
mechanics does not offer a model for the quantum object.

\medskip

\begin{center}
\textbf{\underline{ANISOTROPIC VELOCITY OF LIGHT}}
\end{center}

In order to determine the expression for the anisotropic speed of light in
the Earth's vicinity let us consider three points $A$, $B$ and $C$ on the $x$
axis along the radial direction (when the origin of the Cartesian
coordinates coincides with the Earth's center all coordinate axes $x,$ $y$
and $z$ have radial directions). Light signals originate from point $B$ and
reach point $A$ lying above $B$ at a distance $r$ and point $C$ situated
bellow $B$ at the same distance $r$. To determine the speed of light at $A$
an $C$ as seen from $B$ we need the ratios of the length intervals at $A$, $%
B $ an $C$. In Cartesian coordinates the interval in a gravitational field is%
$^{15}$:

\begin{eqnarray*}
ds^{2} &=&g_{\mu \nu }dx^{\mu }dx^{\nu }=\left( 1-\frac{2GM}{c^{2}R}\right)
c^{2}dt^{2} \\
&&-\left( 1+\frac{2GM}{c^{2}R}\right) \left( dx^{2}+dy^{2}+dz^{2}\right) ,
\end{eqnarray*}
where $M$ is the mass of the gravitating body (in this case the Earth), $R$
is the distance from the body's center to the point for which the interval
is written, $G$ is the gravitational constant and $c$ is the velocity of
light. Then the length interval $dX_{B}$ at $B$ in the $x$ direction is,

\[
dX_{B}\equiv ds_{B}=\sqrt{-g_{11}}dx\approx \left( 1+\frac{GM}{c^{2}R_{B}}%
\right) dx.
\]
At $A$ and $C$ the length intervals $dX_{A}$ and $dX_{C}$ are:

\[
dX_{A}\equiv ds_{A}\approx \left( 1+\frac{GM}{c^{2}R_{A}}\right) dx
\]

\[
dX_{C}\equiv ds_{C}\approx \left( 1+\frac{GM}{c^{2}R_{C}}\right) dx
\]
Then the ratio of the lengths at $A$ and $B$ is:

\[
\frac{dX_{A}}{dX_{B}}=\left( 1+\frac{GM}{c^{2}R_{A}}\right) /\left( 1+\frac{%
GM}{c^{2}R_{B}}\right) \hspace{0.7in}
\]

\begin{equation}
\approx \left( 1+\frac{GM}{c^{2}R_{A}}-\frac{GM}{c^{2}R_{B}}\right) \approx
1-\frac{gr}{c^{2}}  \label{2ratio}
\end{equation}
since $R_{A}=$ $R_{B}+r$ and $GM/R_{B}^{2}=g$, where $g$ is the
gravitational acceleration. The ratio of the lengths at $C$ and $B$ is
analogously

\[
\frac{dX_{C}}{dX_{B}}=\left( 1+\frac{GM}{c^{2}R_{C}}\right) /\left( 1+\frac{%
GM}{c^{2}R_{B}}\right) \hspace{0.7in}
\]

\begin{equation}
\approx \left( 1+\frac{GM}{c^{2}R_{C}}-\frac{GM}{c^{2}R_{B}}\right) \approx
1+\frac{gr}{c^{2}}  \label{3ratio}
\end{equation}
since $R_{C}=R_{B}-r$.

In order to calculate the speed of light at $A$ as seen from $B$ we are
interested in seeing how much of $B$'s proper time $d\tau _{B}$ (measured at
$B$) it will take for the light to travel the distance $dX_{A}$ (at $A$),
which is the proper length of $A$. Two observers at $A$ and $B$ agree that
the distance at $A$ has the magnitude of $dX_{A}$. Using (\ref{2ratio}) we
have for the speed of light at $A$ as seen from $B$%
\begin{eqnarray*}
c_{A} &\equiv &\frac{dX_{A}}{d\tau _{B}}=\frac{dX_{B}\left( 1-\frac{gr}{%
c^{2}}\right) }{d\tau _{B}} \\
&=&c\left( 1-\frac{gr}{c^{2}}\right)
\end{eqnarray*}
since the local speed of light $dX_{B}/d\tau _{B}=c$. Similarly from (\ref
{3ratio}) the speed of light at $C$ as seen from $B$ is
\begin{eqnarray*}
c_{C} &\equiv &\frac{dX_{C}}{d\tau _{B}}=\frac{dX_{B}\left( 1+\frac{gr}{%
c^{2}}\right) }{d\tau _{B}} \\
&=&c\left( 1+\frac{gr}{c^{2}}\right) .
\end{eqnarray*}
In vector notation the anisotropic speed of light in a gravitational field
at a point at a distance $r$ from $B$ as seen from $B$ is:

\[
c^{g}=c\left( 1+\frac{\mathbf{g}\cdot \mathbf{r}}{c^{2}}\right)
\]
The average velocity of light between the source point and the observation
point is:

\begin{equation}
\bar{c}^{g}=c\left( 1+\frac{\mathbf{g}\cdot \mathbf{r}}{2c^{2}} \right)
\label{4c_av}
\end{equation}

Here we consider only small distances for which $\mathbf{g}\cdot \mathbf{r/}%
2c^{2}$ $\ll $ $1.$ This restriction makes it possible for the principle of
equivalence to be applied and to relate results in a reference frame at rest
on the Earth's surface and a reference frame moving with an acceleration $%
\mathbf{a=-g}$.

\medskip

\begin{center}
\textbf{\underline{GRAVITATIONAL ATTRACTION}\\[0pt]
\underline{WITHOUT A GRAVITATIONAL FIELD}}
\end{center}

\medskip

To demonstrate that the anisotropic speed of light in the Earth's vicinity
fully accounts for the gravitational properties of an electron, as discussed
in the introduction, let us first consider a stationary electron in a
non-inertial frame $N^{g}$ at rest on the Earth's surface. The electron's
potential and electric field are distorted due to the anisotropic velocity
of light (\ref{4c_av}). In order to calculate the force of repulsion between
two charge elements $de$ and $de_{1}$ of a non-inertial electron (at rest in
$N^{g}$) we have to find the potential of a charge element $de$. The
anisotropic speed of light (\ref{4c_av}) leads to two changes in the scalar
potential (\ref{pot_in}) of an inertial charge element $de$:
\begin{equation}
d\varphi \left( r,t\right) =\frac{de}{4\pi \epsilon _{o}r}=\frac{\rho dV}{%
4\pi \epsilon _{o}r}{,}  \label{pot_in}
\end{equation}
where $\rho $ is the charge density and $dV$ is the volume of the charge
element. \noindent First, $r$, determined as $r=ct$ (where $t$ is the time
it takes for an electromagnetic signal to travel from the charge element to
the point at which the potential is determined), will have the form $r^{g}=%
\bar{c}^{g}t$ in $N^{g}$. Assuming $\mathbf{g}\cdot \mathbf{r/}2c^{2}$ $\ll $
$1$ we can write:
\begin{equation}
\left( r^{g}\right) ^{-1}\approx r^{-1}\left( 1-\frac{\mathbf{g}\cdot
\mathbf{r}}{2c^{2}}\right) .  \label{r_g}
\end{equation}

The second change in (\ref{pot_in}) is a Li\'{e}nard-Wiechert-like
contribution to the scalar potential which has not been noticed up to now.
It is analogous to the Li\'{e}nard-Wiechert potentials resulting from an
apparently larger dimension of a moving charge (in the direction of its
motion) as viewed by an inertial observer $I$ $^{16-19}$. In $N^{g}$ the
electron is at rest but a volume element of it is apparently different from
the actual volume element $dV$ due to the anisotropic velocity of light (\ref
{4c_av}). The anisotropic volume element (which contains the
Li\'{e}nard-Wiechert-like term) in $N^{g}$ arises from the different average
velocities of electromagnetic signals originating from the rear end and the
front end of the charge element $de$ (with respect to the observation
point), and is given by$^{10}$:

\begin{equation}
dV^{g}=dV\left( 1-\frac{\mathbf{g}\cdot \mathbf{r}}{2c^{2}} \right)
\label{dV_g}
\end{equation}
where $dV$ is the actual volume element (i.e. the volume element determined
when the electron is at rest in an inertial reference frame). Now taking
into account (\ref{r_g}) and (\ref{dV_g}) the scalar potential of a charge
element of the electron becomes

\[
d\varphi ^{g}=\frac{1}{4\pi \epsilon _{0}}\frac{\rho dV^{g}}{r^{g}}=\frac{1}{%
4\pi \epsilon _{0}}\frac{\rho dV}{r}\left( 1-\frac{\mathbf{g}\cdot \mathbf{r}%
}{2c^{2}}\right) ^{2}
\]
or if we keep only the terms proportional to $c^{-2}$

\begin{equation}
d\varphi ^{g}=\frac{\rho }{4\pi \epsilon _{0}r}\left( 1-\frac{\mathbf{g}%
\cdot \mathbf{r}}{c^{2}}\right) dV.  \label{pot_g}
\end{equation}
The potential (\ref{pot_g}) of a charge element $de^{g}$ of a non-inertial
electron contains a Li\'{e}nard-Wiechert-like term (the expression in the
brackets). The electric field of the charge element $de^{g}=\rho dV^{g}$ in $%
N^{g}$ can be directly calculated by using only the scalar potential (\ref
{pot_g}):
\[
d\mathbf{E}^{g}=-\nabla d\varphi ^{g}=\frac{1}{4\pi \epsilon _{o}}\left(
\frac{\mathbf{n}}{r^{2}}-\frac{\mathbf{g\cdot n}}{c^{2}r}\mathbf{n}+\frac{1}{%
c^{2}r}\mathbf{g}\right) \rho dV
\]

\noindent and the field of the electron is
\begin{equation}
\mathbf{E}^{g}=\frac{1}{4\pi \epsilon _{o}}\int \left( \frac{\mathbf{n}}{%
r^{2}}-\frac{\mathbf{g\cdot n}}{c^{2}r}\mathbf{n}+\frac{1}{c^{2}r}\mathbf{g}%
\right) \rho dV{.}  \label{E_g}
\end{equation}

The self-force with which the electron's field interacts with another
element $\rho dV_{1}^{g}$ of the electron charge is
\begin{eqnarray*}
d\mathbf{F}_{self}^{g} &=&\rho dV_{1}^{g}\mathbf{E}^{g} \\
&=&\frac{1}{4\pi \epsilon _{o}}\int \left( \frac{\mathbf{n}}{r^{2}}-\frac{%
\mathbf{g}\cdot \mathbf{n}}{c^{2}r}\mathbf{n}+\frac{1}{c^{2}r}\mathbf{g}%
\right) \\
&& \\
&&\times \rho ^{2}dVdV_{1}^{g}{.}
\end{eqnarray*}

The resultant self-force with which the electron acts upon itself is:
\begin{eqnarray*}
\mathbf{F}_{self}^{g} &=&\frac{1}{4\pi \epsilon _{o}}\int \int \left( \frac{%
\mathbf{n}}{r^{2}}-\frac{\mathbf{g}\cdot \mathbf{n}}{c^{2}r}\mathbf{n}+\frac{%
1}{c^{2}r}\mathbf{g}\right) \\
&& \\
&&\times \rho ^{2}dVdV_{1}^{g}{,}
\end{eqnarray*}

\noindent which after taking into account the explicit form (\ref{dV_g}) of $%
dV_{1}^{g}$ becomes

\begin{eqnarray}
\mathbf{F}_{self}^{g} &=&\frac{1}{4\pi \epsilon _{o}}\int \int \left( \frac{%
\mathbf{n}}{r^{2}}-\frac{\mathbf{g}\cdot \mathbf{n}}{c^{2}r}\mathbf{n}+\frac{%
1}{c^{2}r}\mathbf{g}\right)  \nonumber \\
&&  \nonumber \\
&&\times \left( 1-\frac{\mathbf{g}\cdot \mathbf{r}}{2c^{2}}\right) \rho
^{2}dVdV_{1}{.}  \label{F_g}
\end{eqnarray}

Assuming a spherically symmetric distribution$^{1,\ 2}$ of the electron
charge and following the standard procedure of calculating the self-force$%
^{20}$ we get:

\begin{equation}
\mathbf{F}_{self}^{g}=\frac{U}{c^{2}}\mathbf{g},  \label{F_g1}
\end{equation}
where

\[
U=\frac{1}{8\pi \epsilon _{o}}\int \int \frac{\rho ^{2}}{r}dVdV_{1}
\]

\noindent is the electron's electrostatic energy. As $U/c^{2}$ is the mass
''attached'' to the field of an electron, i.e. its electromagnetic mass, (%
\ref{F_g1}) obtains the form:

\begin{equation}
\mathbf{F}_{self}^{g}=m^{g}\mathbf{g,}  \label{F=mg}
\end{equation}

\noindent where $m^{g}$ here is interpreted as the electron's passive
gravitational mass. The self-force $\mathbf{F}_{self}^{g}$ which acts upon
an electron on account of its own distorted field is directed parallel to $%
\mathbf{g}$ and resists its acceleration arising from the fact that the
electron (at rest on the Earth's surface) is prevented from falling, i.e.
from moving by inertia. This force is traditionally called a gravitational
force but as we have seen $\mathbf{F}_{self}^{g}$ in (\ref{F=mg}) is purely
electromagnetic in origin. This result explains why general relativity
predicts that there is no gravitational force. The spacetime anisotropy
around the Earth is sufficient to account for the force an electron on the
Earth's surface is subjected to; this force, however, is not gravitational
but electromagnetic.

The famous factor of $4/3$ in the electromagnetic mass of the electron does
not appear in (\ref{F=mg}). The reason is that in (\ref{F_g}) we have used
the correct volume element $dV_{1}^{g}=\left( 1-\frac{\mathbf{g}\cdot
\mathbf{r}}{2c^{2}}\right) dV_{1}$. This apparent change of the volume
element originates from the anisotropic speed of light in a non-inertial
frame and taking it into account naturally removes the $4/3$-factor without
resorting to the Poincar\'{e} stresses (designed to explain the stability of
the electron). Since its origin a century ago the electromagnetic mass
theory of the electron has not been able to explain why the electron is
stable (what holds its charge together). This failure has been used as
evidence against regarding its entire mass as electromagnetic; it has been
assumed that part of the electron mass originates from forces holding the
electron charge together (known as Poincar\'{e} stresses). However, the
problem of stability of the electron cannot be adequately addressed until a
quantum-mechanical model of the electron structure is obtained. On the other
hand, this problem can be successfully avoided in the case of the
electromagnetic mass derived from the expression for the momentum of the
electron's electromagnetic field$^{8,\ 9}$. The stability problem does not
interfere, as we have seen, with the derivation of the expression for the
self-force containing the electromagnetic mass either. This hints that
perhaps there is no real problem with the stability of the electron (as a
future quantum mechanical model of the electron itself may find); if there
were one it would inevitably emerge in the calculation of the self-force.

General relativity describes an electron falling in a gravitational field by
a geodesic worldline. It implies that it moves by inertia and its Coulomb
field should not be distorted which means that there should not exist any
self-force acting on the electron. The electron's Coulomb field is not
distorted as viewed by an inertial observer $I$ falling with the electron.
In order to obtain the electric field of an accelerated electron falling in
the Earth's gravitational field ($\mathbf{a}=\mathbf{g}$) with respect to a
non-inertial observer (at rest in $N^{g}$) we cannot use the
Li\'{e}nard-Wiechert potentials in $N^{g}$ since they are valid only in an
inertial reference frame ($N^{g}$ is a non-inertial frame). Due to the
anisotropic speed of light (\ref{4c_av}) in $N^{g}$ they must include the
Li\'{e}nard-Wiechert-like term, contained in the potential (\ref{pot_g}):
\begin{equation}
\varphi ^{g}\left( r,t\right) =\frac{e}{4\pi \epsilon _{o}}\frac{1}{r-%
\mathbf{v\cdot r}/c}\left( 1-\frac{\mathbf{g}\cdot \mathbf{r}}{c^{2}}\right)
\label{LWs_g}
\end{equation}

\begin{equation}
\mathbf{A}^{g}\left( r,t\right) =\frac{e}{4\pi \epsilon _{o}c^{2}}\frac{%
\mathbf{v}}{r-\mathbf{v\cdot r}/c}\left( 1-\frac{\mathbf{g}\cdot \mathbf{r}}{%
c^{2}}\right) .  \label{LWv_g}
\end{equation}
The electric field of an electron falling in $N^{g}$ (and considered
instantaneously at rest\footnote{%
The only reason for considering the instantaneous electric field is to
separate the deformation of the electric field due to the Lorentz
contraction from the distortion caused by the acceleration.} in $N^{g}$)
obtained from (\ref{LWs_g}) and (\ref{LWv_g}) is:

\[
\mathbf{E}=-\nabla \varphi ^{g}-\frac{\partial \mathbf{A}^{g}}{\partial t}%
{ \hspace{0.13in} \hspace{0.3in} \hspace{0.1in} \hspace{0.4in}}
\]

\[
=\frac{e}{4\pi \epsilon _{o}}\left( \frac{\mathbf{n}}{r^{2}}+\frac{\mathbf{%
g\cdot n}}{c^{2}r}\mathbf{n}-\frac{1}{c^{2}r}\mathbf{g}\right)
\]

\[
+\frac{e}{4\pi \epsilon _{o}}\left( -\frac{\mathbf{g}\cdot \mathbf{n}}{c^{2}r%
}\mathbf{n}+\frac{1}{c^{2}r}\mathbf{g}\right) .
\]

\noindent In such a way, the electric field of a falling electron in the
reference frame $N^{g}$ proves to be identical with the field of an inertial
electron determined in its rest frame:
\begin{equation}
\mathbf{E}=\frac{e}{4\pi \epsilon _{o}}\frac{\mathbf{n}}{r^{2}}.
\label{E_in}
\end{equation}

It follows from (\ref{E_in}) that both an inertial observer (falling with
the electron) and a non-inertial observer (at rest in $N^{g}$) detect a
Coulomb field of the electron falling in $N^{g}$. In other words, while the
electron is falling in the Earth's gravitational field its electric field at
any instant is the Coulomb field which means that no force is acting on the
electron, i.e. there is no resistance to its accelerated motion. This result
sheds light on the fact that in general relativity the motion of a body
falling toward a gravitating center is regarded as inertial (non-resistant)
and is described by a geodesic worldline. Now we are in a position to answer
the question why an electron is falling in a gravitational field and no
force is causing its acceleration. As (\ref{E_in}) shows, the only way for
an electron to compensate the anisotropy in the propagation of the
electromagnetic signals (responsible for the repulsion force each volume
element of it is subjected to) and to keep its electric field not distorted
is to fall with an acceleration $\mathbf{g}$. If the electron is prevented
from falling its electric field distorts due to the anisotropic speed of
light and the self-force (\ref{F=mg}) appears. It tries to force the
electron to move (fall) in such a way that its field becomes the Coulomb
field and as a result the self-force disappears.

We have, on the one hand, the result (\ref{E_in}) which demonstrates that a
Coulomb field is associated with a falling electron by \emph{both} an
inertial observer $I$ (falling with the electron) and a non-inertial
observer at rest in $N^{g}$. On the other hand, comparing the electric field
(\ref{E_g}) of an electron at rest in $N^{g}$(determined in $N^{g}$) and its
field$^{19,\ 20}$ determined in $I$, in which the electron is
instantaneously at rest (having an acceleration $\mathbf{a=-g}$) shows that
for both an observer in $I$ and an observer in $N^{g}$ the electron's field
is equally distorted. This result reveals that there exists a unique
connection between the shape of the electric field of an electron and its
inertial state: if an electron is represented by a geodesic worldline (which
means that it moves by inertia) its field is the Coulomb field - both an
inertial observer $I$ and a non-inertial observer $N^{g}$ detect the same
(Coulomb) field; if the worldline of an electron is not geodesic (meaning
that the electron does not move by inertia), its electric field is deformed
- both $I$ and $N^{g}$ observe the same distorted electric field. Stated
another way, the inertial state of an electron is absolute and for this
reason the shape of its electric field is also absolute (the same for both
an inertial observer and a non-inertial observer).

\medskip

\begin{center}
\textbf{\underline{CONCLUSIONS}\\[0pt]
}
\end{center}

\medskip

The consequence of general relativity that the velocity of light around
massive bodies is anisotropic and the classical electromagnetic mass theory
reveal that gravitation is electromagnetic in origin:

(i) Due to the anisotropic speed of light the electric field of an electron
on the Earth's surface is distorted which gives rise to an electric
self-force trying to force the electron to move downwards. The self-force,
being proportional to $\mathbf{g}$ with the coefficient of proportionality
representing the mass that corresponds to the energy stored in the
electron's electric field, is equal to what is traditionally called a
gravitational force. In such a way, the nature of the force acting on a body
on the Earth's surface is electromagnetic (since the body's constituents are
all charged particles at the most fundamental level). This means that a
body's passive gravitational mass is electromagnetic in origin too.

(ii) An electron is falling toward the Earth with an acceleration $\mathbf{g}
$ in order to compensate the anisotropy in the propagation of
electromagnetic signals (with which the different charged elements of the
electron repel one another) which ensures that its electric field does not
distort. The electron is not subjected to any self-force only if its
electric field is the Coulomb field. If the electron is prevented from
falling the compensation of the anisotropy of the speed of light is not
possible any more and its electric field gets deformed which gives rise to a
self-force pulling the electron downwards. This mechanism explains why all
bodies fall toward the Earth with the \emph{same} acceleration - each of
their elementary charged constituents is falling with an acceleration $%
\mathbf{g}$ in order to prevent its electric field from being distorted.

It is believed that the anisotropy of the speed of light around a massive
body is caused by the curvature of spacetime around the body (i.e. by the
body's gravitational field). The spacetime curvature itself originates from
the body's active gravitational mass. As we have seen, however, the mass of
a body is electromagnetic in origin - this is the mass that corresponds to
the energy stored in the electric fields of the body's elementary charged
constituents. As there is no mass but only charges (and their fields), it
follows that the anisotropy of the speed of light around a body is caused by
the body's charges and their fields. In such a way, the spacetime curvature
proves to be a spacetime anisotropy. And since charges and electromagnetic
fields are in principle controllable, the anisotropy of spacetime and
spacetime geometry itself are in principle controllable as well.

The theoretical possibility to manipulate the spacetime geometry constitutes
a radically new approach to the understanding of how a body can be
propelled. There is no need for any force to be applied to a body itself in
order to propel it. Instead, the geometry of spacetime can be subjected to a
local manipulation through the application of the sources of spacetime
anisotropy - charges and electromagnetic fields. By changing the anisotropy
of spacetime we are in fact changing the spacetime geometry. This in turn
leads to a change of the shape of a body's geodesic worldline and ultimately
to a change of a body's state of motion. In other words, a body can be
propelled without being subjected to any direct force. This paper has
demonstrated that this is possible at least in principle. How it can be done
is the subject of ongoing work.

\medskip

\begin{center}
\textbf{\underline{ACKNOWLEDGMENTS}\\[0pt]
}
\end{center}

\medskip

I would like to acknowledge helpful discussions and correspondence with Dr.
B. Haisch and Prof. A. Rueda.

\medskip

\begin{center}
\textbf{\underline{REFERENCES}}
\end{center}

[1] Abraham, M. (1950) \emph{The Classical Theory of Electricity and
Magnetism}\textit{,} 2nd ed. London, Blackie. 

[2] Lorentz, H. A. (1952) \emph{Theory of Electrons}, 2nd ed. New York,
Dover. 

[3] Fermi, E. (1922) \emph{Phys. Zeits.} \textbf{23}, 340.

[4] Mandel, H. (1926) \emph{Z. Physik} \textbf{39}, 40. 

[5] Wilson, W. (1936) \emph{Proc. Phys. Soc.} \textbf{48}, 736.

[6] Pryce, M. H. L. (1938) \emph{Proc. Roy. Soc. A}\textbf{\ 168}, 389.

[7] Kwal, B. (1949) \emph{J. Phys. Rad.} \textbf{10}, 103.

[8] Rohrlich, F. (1960) \emph{Am. J. Phys.} \textbf{28}, 63.

[9] Rohrlich, F. (1990)\emph{\ Classical Charged Particles,} New York,
Addison-Wesley. 

[10] Petkov, V. \emph{Physics Letters A}, submitted. 

[11] Haisch, B., Rueda, A. and Puthoff, H. E. (1994) \emph{Phys. Rev. A}
\textbf{\ 49}, 678. 

[12] Puthoff, H. E. (1989) \emph{Phys. Rev. A}\textbf{\ 39}, 2333.

[13] Rueda, A. and Haisch, B. (1998) \emph{Phys. Letts. A} in press.

[14] Rueda, A. and Haisch, B. (1998) \emph{Found. Phys.} in press.

[15] Ohanian, H. and Ruffini, R. (1994) \emph{Gravitation and Spacetime},
2nd ed., New York, London, W. W. Norton, p. 179. 

[16] Feynman, R. P., Leighton, R. B. and Sands, M. (1964) \emph{The Feynman
Lectures on Physics, }Vol. 2, New York, Addison-Wesley. 

[17] Panofsky, W. K. H. and Phillips, M. (1962) \emph{Classical Electricity
and Magnetism,} 2nd ed., Massachusetts, London, Addison-Wesley.

[18] Schwartz, M. (1972) \emph{Principles of Electrodynamics}, New York,
Dover. 

[19] Griffiths, D. J. (1989) \emph{Introduction to Electrodynamics}\textit{,}
2nd ed., New Jersey, Prentice Hall. 

[20] Podolsky, B. and Kunz, K. S. (1969) \emph{Fundamentals of
Electrodynamics,} New York, Marcel Dekker. 

\end{document}